# Genetic Algorithm Based Resource Minimization in Network Code Based Peer-to-Peer Network

M. Anandaraj[*]
*Department of Information Technology,*
*PSNA College of Engineering and Technology, Dindigul, Tamilnadu, India.*
*ananddgl@yahoo.com*

K. Selvaraj
*Department of Information Technology,*
*PSNA College of Engineering and Technology, Dindigul, Tamilnadu, India.*
*selme04@gmail.com*

P. Ganeshkumar
*College of Computer and Information Science,*
*Al-Imam Mohammad Ibn Saud Islamic University, Saudi Arabia*
drpganeshkumar@gmail.com

K. Rajkumar
*Department of Information Technology,*
*PSNA College of Engineering and Technology, Dindigul, Tamilnadu, India..*
*rajkumardgl@psnacet.edu.in*

K. Sriram
*Department of Biomedical Engineering,*
*PSNA College of Engineering and Technology, Dindigul, Tamilnadu, India..*
*sriram@psnacet.edu.in*

Block scheduling is difficult to implement in P2P network since there is no central coordinator. This problem can be solved by employing network coding technique which allows intermediate nodes to perform the coding operation instead of store and forward the received data. There is a general assumption in this area of research so far that a target download rate is always attainable at every peer as long as coding operation is performed at all the nodes in the network. An interesting study is made that a maximum download rate can be attained by performing the coding operation at relatively small portion of the network. The problem of finding the minimal set of node to perform the coding operation and links to carry the coded data is called as a network code minimization problem (NCMP). It is proved to be NP hard problem. It can be solved using genetic algorithm (GA) because GA can be used to solve the diverse NP hard problem. A new NCMP model which considers both minimize the resources needed to perform coding operation and dynamic change in network topology due to disconnection is proposed. Based on this new NCMP model, an effective and novel GA is proposed by implementing problem specific GA operators into the evolutionary process. There is an attempt to implement the different compositions and several options of GA elements which worked well in many other problems and pick the one that works best for this resource minimization problem. Our simulation results prove that the proposed system outperforms the random selection and coding at all possible node mechanisms in terms of both download time and system throughput.

## 1. Introduction and Related Work

Distribution of content over the internet means to deliver digital content such as text and multimedia content, software and live streaming audio and video to a large group of users. Traditionally it can be done with the help of client-server system in which client sends request to the server and receive requested file as a response from the server. To improve the performance of the system with respect to response and process time, a content delivery network has been introduced. It locates a collection of servers in geographically distributed manner. Commercial CDNs such as Digital Island and Akamai offer this service for several popular commercial sites. Peer-to-Peer (P2P) network is an alternative to this traditional client-server system for distributing the large scale multimedia content[1]. It is a distributed and cooperative architecture. In a P2P network model, each peer receives assistance from other peers for further distribution of content to others. These networks are scalable one. The new nodes join the network not only bring new demand and also additional computing power and bandwidth to the network. There is no longer a single point of failure in the network. Even P2P network makes the network resources sufficient; it still faces the challenge of how the available resources are used optimally without having any central coordinator with low complexity. These networks are highly dynamic such that nodes joining and leaving the network at any time and node has only local topological information about the network. BitTorrent is a popularly known P2P system. It utilizes swarming techniques to simultaneous download of different segments of a same file among peers. Downloading file by collecting its segment can be modeled by classic coupon collector problem. In this, probability of gathering fragment falls rapidly with the number of those already collected. As the number of nodes increases, it is difficult to schedule the distribution of packet to other nodes optimally. Download local rarest segment first is a possible solution for the above said problem. But, most often such segment fails to match those that are globally rarest one. This leads to slower download rate and increase of failure rate.

Network coding may be used to P2P network in practice to resolve the block scheduling problem in a simple and effective manner. It leads to shorter file downloading time. Furthermore, it offers robustness against peer dynamics and makes each packet equally important. Peers are computing devices which are able to code the incoming packets by applying network coding in software. It is not needed to revise the existing switches and routers in the Internet. Network coding technique attracted considerable research interest since the publication of [2], which described its effectiveness for multicasting in wire line network. It is now recognized that network coding provides lots of benefits to both wire and wireless network in terms of better throughput, balanced network load, security, robustness against link failures and so on[3-5]. Several research works in recent years proved that network coding can also be used to improve the efficiency of content distribution such as application level multicast, P2P streaming applications and reliable large scale multimedia content distribution[6-10].

It is found that conventional network optimization intends to exploit information flow by using as much link capacity as possible [11- 12]. At the same time network coding begins with the assumption that full link capacity utilization has been achieved wherever possible and then tries to further enhance the network throughput at the receiver by carrying out coding at intermediate nodes [13]. Even though network coding is performed at all the nodes in some relevant literature, a remarkable observation is that a maximum download rate can be achieved by performing the coding operation at relatively small portion of the network [14]. To date, though random network coding emerges as a promising alternative to the traditional methods in theoretical perspective, it's benefit in real-world content delivery systems have not been examined fully in terms of number of significant performance metrics, such as average download completion times and peer dynamics[15]. Hence, a question is raised: which nodes in the network have to carry out coding operations or how does one use most of network capacity at a nominal cost with respect to network coding resources?. To response this question, a minimal set of nodes need to be selected to carry out the coding operation, and this is proved to be NP-hard problem. In this article, the above stated problem of minimizing network coding resources is called as Network Coding Minimization Problem (NCMP). Several attempts have been made already using different techniques to solve this problem.

This article is organized as follows. Section 2 describes the literature study carried out to find the solution for proposed research problems. Section 3 illustrates the network code based P2P content distribution system. In section 4, the proposed system model is discussed. Section 5 presents the design of proposed GA. In section 6, the main experimental results and analysis are provided and conclude the proposed work in section 7.

## 2. Related Work

A significant advantage of network coding in P2P network is the robustness to peer departures. Without network coding, there is a possibility that some of data blocks are lost, because of departure of the peers as well as server who have these blocks of data. In this inopportune event, the original content cannot be reconstructed by the remaining peers. By using network coding, the risk of losing certain data blocks is no longer a major concern. Since data blocks are combined together, each of coded blocks is spreading to a large number of coded blocks in the system. Two minimal approaches have been reported in [16,17] which determine the minimal set of nodes to attain a maximum possible target rate. The approaches in [16,17] find the minimal set of nodes to perform the coding operation by removing connections in a greedy manner. A linear programming technique was stated in [17] to optimize the different resources utilized for network coding. Its best possible formulations involved number of variables and constraints that increase exponentially with the number of users increased in the network. This makes it hard to apply the formulations in case of a large number of nodes, even at the price of forfeited optimality. Genetic algorithms (GAs) have been proved to solve the various NP-hard problems [18,19], including optimization of network resources and resource assignment problem[20,21]. The first effort to apply GA to network coding was performed by Kim et al. [22]. Their work was extended from acyclic networks to cyclic networks and also from centralized one to decentralized one further in [23]. The genetic representation was the focus in their later work [24], followed by the suggestion of a distributed algorithm to enhance GA computational efficiency [26]. Instead of tackling this NP-hard problem, our focus is to find optimal solution quickly. Finding a good order of path traversal out of exponentially many possible sequences is vital to the quality of the solutions by the above mentioned minimal approaches. The problem of finding where to perform the coding operation requires selection out of a many number of choices. Our method manages a set of candidate solutions of a suitably less in number and enhancing solutions in an evolutionary manner. The following table 1 represents comparative analysis of some of the works carried out related to the proposed work.

| Author | Proposed work | Analyze |
|---|---|---|
| Arne Vater, et al[27] | Tree-structured network-coded P2P system. | A tree structure contains leaves, internal nodes (which include leaves or other internal nodes as children), and a root node. All nodes except leave must contain at most two child nodes. The leaves of the tree are novel data blocks multiplied by a coefficient. In our proposed work, GA is used to find the minimum set of nodes to perform the coding operations. ` |
| Guanjun et al. [28] | sparse coding scheme | Instead of encoding operation performed at all the pieces, only a few pieces are chosen randomly to be encoded aiming to alleviate encoding and decoding computational complexities in sparse coding scheme. Same kind of approach is used in our proposed work where instead of selecting the pieces in random, node to perform the network coding is chosen by using GA approach in our proposed work. |
| Langberg M et al. and Bhattad et al. | Greedy approach | Two minimal approaches have been reported in [14,15] which find the minimal set of nodes to reach the maximum possible target rate. The approaches in [14,15] find the minimal set of nodes to perform the coding operation by removing connections in a greedy manner. Instead of greedy approach to minimize the coding operation, nodes are selected to carry out the |
| M Ananda raj et al. | Bit Torrent | Bit Torrent is more notable and prominent P2P network over Avalanche, NC-infused from Microsoft is the answer. Computational complexity is perhaps the major reason Network Code based P2P networks are not popular. The target computer specifications may differ from P2P. Hence, computers which have less computational resource may suffer the performance of said peer in the network, because it would take longer time to encode and decode data blocks. Keep the complexity of NC minimum is mandatory to make the NC to become popular. Hence in our proposed work, GA based model is proposed. |

**Table 1**. Comparative Analysis

This paper aims to propose an efficient GA based approach to solve the NCMP problem and our contributions in this paper are twofold.

- A problem specific genetic algorithm based approach to decrease the number of nodes to perform the coding operation in order to decrease the computational complexity involved in network code based P2P content distribution at the same time realize the benefits of the network coding technique is proposed.

- GA's performance heavily relies on the details of its elements such as population initialization, crossover operation, mutation operation, etc. Theory to accurately calculate which combination of such elements is well suited to a specific problem is not yet available. Hence, we look at different composition and several options of those elements which work well in many other problems and pick the one that works best for our node selection problem.

## 3. Network Code Based P2P Content Distribution

Randomized network coding was first introduced in information coding theory [4] and it has been used in P2P content distribution to enhance system performance in terms of download time and throughput [5]. Each peer can linearly combine the blocks it holds using random coefficients and sends the generated coded blocks to its downstream peers in randomized network coding settings. This technique simplifies P2P protocol design by making each packet equally important [26]. Hence, data scheduling is easier and simple. There are two P2P content distribution system is demonstrated in **Fig. 1(a)** and **Fig. 1(b)**. The first one does not use coding at all, whereas the second one uses network coding across all existing blocks. It is much harder for a non-coding protocol to guarantee a uniform distribution of all blocks in dynamic P2P networks. In **Fig. 1(a)**, if peer N2 leaves the network, any other peers cannot able to obtain the block four since it is available only with peer N2. In contrast, when randomized network coding is used (**Fig. 1(b)**), all coded packets P are equally important with high probability. Even if both peers N4 and N2 leave the network, the content is still decodable for both peer N1 and N3 since the peer N1 and N3 jointly have five coded blocks.

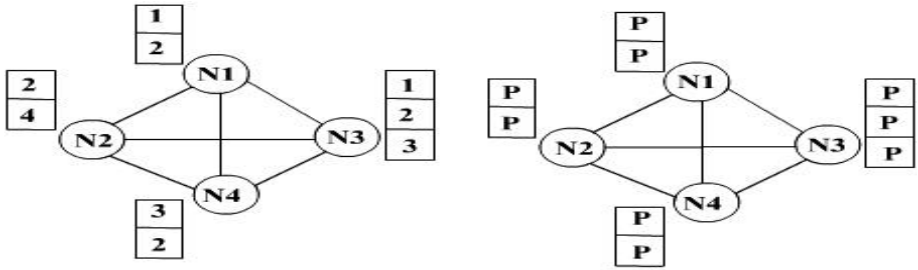

**Figure 1 a.**                    **Figure 1 b.**

**Fig. 1.** P2P content distribution with and without network coding

Network coding usage leads to reduce the download time since peers are not going to be delayed in finding rarest blocks in the network [26]. The difficulty of locating rarest blocks in a non-coding protocol does not present in network code based approach because of all coded blocks is equally innovative and useful to other peers in the network. The implementation of Network coding in dynamic environments such as P2P network is a recent and thrust research area. Random linear network coding (RLNC) was proved to be promising techniques to cope up with network topology changes since it does not require entire network topological information. Hence it can be used in P2P network because of its dynamic nature and difficulty to obtain the entire topological information. In these RLNC mechanisms, the coding resource minimization is not a major concern and each node has to exchange its coding weights throughout the entire network. This exchange leads to the under utilization of network capacity so it is necessary to investigate how to enhance the performance of network coding based on the entire network topology at the same time the advantages of RLNC need to be acknowledged. Robustness against dynamic change in network topology will be the main concern. The dynamic network coding model introduced in this article considers the robustness to some extent when compared to previous works.

## 4. System Model

A P2P large scale content distribution problem is considered where single source has the entire content and send it to many nodes in the network. The nodes are called as peers. Each peer can create and preserve overlay links to some other peers in the network at random, over which content are distributed. The content is divided into S equal size blocks, which are distributed in the system in parallel. At the beginning, each peer doesn't have any part of the content. Most of the existing network coding based P2P content distribution protocols use the following segment based method. Before the distribution, the original content is divided and grouped into segments, each including B blocks of size S bytes. The encoding operations are done within each segment. Each segment is represented as a matrix (M) of size B × S, with rows being the B blocks, and columns being the bytes of each block. The coding operation generates a linear combination of the original blocks in this segment by C = R X M, where R is a B × B matrix consist of random linear coefficients taken from Galois field $GF(2^q)$. The coded blocks such as rows in C, together with the coding coefficients such as rows in R, are coded packet and delivered to other peers in the network.

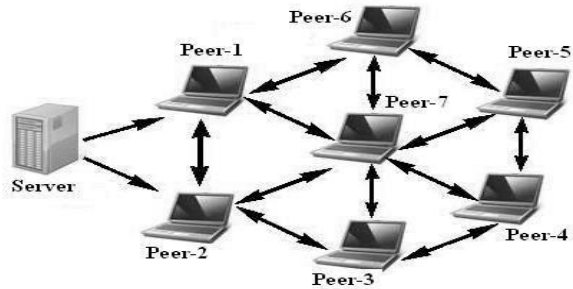

**Fig. 2.** System Model

**Fig. 2** shows the proposed system model. If any peer is selected to perform the coding operation, then it uses random linear network coding (RLNC) [3, 8] to generate new coded blocks from the blocks it has received from other peers. Using RLNC, a coding vector of N coefficients is appended to each coded block to indicate how that coded block is created from the N original blocks. The newly generated coded block together with its encoding vector is then transmitted to the requesting peer. Before requesting a coded block from others, peers need to confirm the encoding vector of that block is linearly independent with what they currently have in their buffer. This avoids downloading of duplicated meaningless coded blocks. Not all the peers in the network to perform the coding operation in selective coding: only some peers perform the coding operation and the others do not. Since coded blocks have been created by randomly combining multiple blocks, the receiving peers choose coded blocks over non coded block. This can improve throughput and simplify the block scheduling of the system. After a peer receives N independent coded blocks (N associated encoding vectors form a full-rank matrix), it can decode to obtain the N original blocks by solving the set of N linear equations using Gaussian elimination method. Network coding does not come without any sacrifice: every node needs to perform the coding operation which incurs computational cost and delay [29, 30]. Our goal is to keep the optimality of network coding at the same time substantially reduce the number of nodes to perform the coding operation. It is also stated as minimize the coding operation without sacrifice the benefits of network coding. If the linearly coded output produced from a peer with multiple incoming links weights all but one incoming data by zero, then efficiently no coding occurs on that link; even if the only nonzero coefficient is not unique one, there is another coding plan that replaces the coefficient as a unique one. Therefore, to identify the peers where coding is not needed, it is necessary to verify at each of the peers with multiple incoming links whether we can limit the peer's outputs to relay on a single input without destroying the achievable target download rate.

### 4.1  Problem Statement

Random linear network coding is used in the proposed system, the coding complexity of the system is described as *O(ECB)* where E is the number of nodes out of N nodes in the system to perform the coding operation, B is the number of original blocks to be distributed, and C is the size of the original content. Since C is fixed for a given content, and B cannot be low value, it is significant to reduce the number of coding nodes to decrease the coding complexity. Nodes in the P2P network form a directed overlay topology and its defined as directed graph $G = \{N, L\}$ where N represents the set of peers or nodes and $L$ is the set of directed overlay links which connects the peers.

The following P2P content distribution based on network coding is used

   • Network topology $G = \{P, L\}$,

   • Source in P with a content of size C to be distributed to all peers in the network, and

   • Number of coding nodes in the network E ($1 \leq E \leq |P|$),

Our problem is to find the optimal node in the network topology where we can perform the coding operation in order to shorten download time and avoid minimize the flooding of useless coded packets. Since our problem is proved as NP-hard [26], we aim at heuristic algorithm which exploits the characteristics of the network topology. GA is a kind of stochastic search algorithm. It functions on a set of possible solutions that are improved sequentially using procedure inspired by biological evolution [30]. It has been used to the various engineering and scientific problems, including multi objective optimization problems in networks [33]. In this paper, there is an attempt to make use of GA to solve the above stated problem.

**GA Notations:**

The following table 2 represents the general GA terms and its descriptions.

| GA Terms | Description |
|---|---|
| Chromosome (String, individual) | Candidate Solution (encoding the solution) |
| Genes (Bits) | Part of candidate solution |
| Locus | Position of gene in the solution |
| Alleles | Value of gene |
| Phenotype | Decoded candidate solution |
| Genotype | Encoded candidate solution |
| Crossover probability | Probability of a pair of chromosomes will be crossed |
| Mutation probability | Probability of a gene on a chromosome will be altered randomly |

**Table 2**. GA Terms and its Description

## 5. Design of Proposed GA

### 5.1. Fitness function formation

An algebraic formulation proposed in [34] for common network coding problem can be used to the situation where network coding is done simply in a small portion of the network. Since our choice in the selection of coding at a node is based on the examination of all of its outgoing and incoming links, and the number of coding links is the exact measure of the amount of computation incurred by coding, our objective is to reduce the number of coding links and coding nodes. Peer with multiple incoming and outgoing links are called as likely to be a coding peer (LCP). We need to find that whether it's possible to reduce the given peer's output to rely on a single input without reducing the achievable download rate. This can be done by verifying the polynomials of a binary matrix $B_x$ which is created based on the coefficients related to the incoming links of each LCP [13]. $B_x$ is feasible only if the coding scheme described by it is good and there is a possibility of attaining target download rate $A_T$ at every peer. It provides enough information to compute the number of coding nodes and links in static network. But the initially calculated possible download rate may not be assured to achieve due to the dynamic nature of P2P network. It is needed to compute the new achievable download rate when there is a change in network topology and then fully optimize the coding plan. Hence $B_x$ is not sufficient to use in the proposed dynamic minimization problem. The problem can be devised mathematically as follows

$$\underset{u(p,q,r)}{Max} \ F, \ p=1,\ldots,m, \ q = 1,\ldots, m_{out}, \ r = 1,\ldots,m_{in}(j) \qquad (1)$$

Subject to the current topology of the network at time t, $G(N(t), L(t), t)$,

$$D_{out} = \sum u(p,q,r)D_{in}(p,r), \ p = 1,\ldots,m_n(p), q = 1,\ldots,m_{out}(p)$$

$$(2)$$

$$F= \begin{cases} a_1 \min(A_T(p) + a_2\,avg(A_T(p)) + a_3/(N_n+1) + a_4/(N_l+1), & \min(A_T(p) < A_T \\ a_1 \min(A_T(p) + a_2\,avg(A_T(p)) + a_5/(N_n+1) + a_6/(N_l+1), & \min(A_T(p) \geq A_T \end{cases} i=1,...,m_n \qquad (3)$$

$$\min(a_{1,} a_{2)} > \max(a_{3,} a_4), \qquad (4)$$

$$\min(a_5, a_6) > \max(a_1,a_{2)} \qquad (5)$$

In the above equation, F is the objective function with $a_i$, i = 1,…,6, as specific coefficients to choose the contribution terms in F (objective function). $m_{in}(p)/m_{out}(p)$ represents the number of incoming and outgoing paths of peer p. The data packet $D_{in}(p,q)$ and $D_{out}(p,q)$ are those on the $q^{th}$ incoming and outgoing links of peer p. The weights u(p,q,r), r = 1,…,$m_{in}(p)$, determine how to combine the $m_{in}(p)$ incoming data of peer p to generate a coded packet for the $q^{th}$ outgoing link of peer p. A sufficient number of finite discrete values for u(p,q,r) can assure that the maximum possible throughput is obtained in terms of download time and system throughput. A set of u(p,q,r) defines how each peer in the network forwards, replicates and/or generate the coded data. Hence, our focus in this paper is to find the optimal value of u(p,q,r) from a finite set. A linear coding scheme determined by u(p,q,r), $A_T(p)$ is the actual download rate achieved at peer p. The optimization of coding scheme, is subject to the network topology defined by G(N(c), L(c), c). The weights u(p,q,r) need to be optimized in the NCMP. The objective function defined, firstly attempts to maximize the overall actually achieved download rate and after the target download rate is attained, then focus of the optimization changes to minimizing the number coding nodes and coding links in the network. The terms $\min(A_T(p))$ and $avg(A_T(p))$ in the equation (3) is used to verify the download rate that is actually attained. We are looking to have a larger value for $avg(A_T(p))$. The term $\min(A_T(p))$ is used to calculate how evenly $A_T(p)$ is optimized. Whilst as reflected by the terms $1/(N_n+1)$ and $1/(N_l+1)$ in which the network coding links and nodes are minimized. The coefficients $a_1$ to $a_6$ satisfy the conditions (4) and (5), play an important role in the changing the focus of optimization from the actually achieved download rate to minimization of network coding resources. The objective function makes it possible to estimate the $A_T(p)$, $N_n$ and $N_l$ in the network.

## 5.2. Representation of candidate solutions

In GA mechanism, a candidate solution $s \in S$ is called as chromosomes. Each chromosome is made up of distinct units called as genes. Each gene controls one or more features of the chromosome. In the original implementation of GA, genes are assumed to be binary numbers. This process is called as a mapping mechanism between the solution space and the chromosomes. This mapping is called as a representation of candidate solution or encoding. The representation of candidate solutions is important for designing efficient and effective evolutionary algorithms in various difficult optimization problems, including the recently emerged coding resource minimization problem concerned to our work. Our problem is similar to ordering problems, such as traveling salesman problem or task ordering problem. There are several methods to represent the candidate solutions for ordering, such as one-dimensional binary code, tree code and sequence and topology encoding such as ST encoding [33]. These representations have low efficiency since they required search space and complexity rise with increase of nodes in the network [35]. Hence we use the permutation encoding method to represent the candidate solutions. With this method of representing the candidate solutions, every chromosome is a string of numbers that denote a place in a sequence.

The proposed chromosome structure based on relative information flow on path record packet flow on this path, path's positions in the network, relative information flow on this path and incoming path's state. This new coding scheme to represent solution must craft it possible to estimate the actually achieved download rate $A_T$ at each peer for objective function. A straight forward approach is to employ the absolute information flow on path to code each solution. It's very difficult to use absolute information flow in the evolutionary approach since it causes serious problem during evolutionary operations. This means that any variation in the information flow on a path caused by evolutionary operations could make the unchanged information flow on some other paths infeasible. Hence it is proposed to record relative information flow in candidate solution instead of absolute information flow. Let consider the integer d, which is interpreted as a certain predefined combination of information on incoming paths of a node. This value is also incorporated into the candidate solution representation.

It is noted that the new chromosome structure relies on topological information of the entire network, which may change often in NCMP. Hence, the proposed GA cannot attain the robust performance defined by RLNC[37][38]. Since the new chromosome structure includes to record whether an incoming path will contribute to a coding occurrence or not, it may deliver robust performance. Hence, when a noncontributing incoming links are broken, it is not necessary to run the optimization again. Because the proposed new structure records exact information flow, it is robust against any change in non contributing incoming paths and also if a contributing path is detached, we can easily verify whether or not there is any non contributing path which can substitute for the broken path [39]. The information flow to a LCP should be spread as much as possible to allow many different options for coding operation so it helps to spread a generation. For a LCP with multiple outgoing paths, there is a high probability that the outgoing paths have different states. The following steps are used to improve up to one gene in a chromosome. By performing the following steps repeatedly, further genes can be enhanced in the chromosome. Hence, the optimal solution can be obtained quickly and it requires less number GA runs. The **Fig 3** represents the proposed GA run to NCMP.

**Algorithm for GA run to NCMP**

| Input   : Set of Peers $N_n$ |
|---|
| Output : Set of LCP $N_l$ |
| Initialization |
|         Find the $A_T(p)$ for each link |
|         Calculate Avg $A_T(p)$ |
| For i=1 to $N_n$ |
|         Generate coefficients related to the incoming links of each LCP for $B_x$ |
|         Generate individuals by verifying outgoing links |
|         Replace the infeasible chromosomes by all-one vectors. |
| Loop |
|         Evaluate the individual by assessing weight of u(p,q,r) |
|         Update LCP by using the NCMP Fig 3. |
|         Apply parameterized uniform crossover with probability 0.08 |
|         Mutate with probability 0.01 |
|         Generate a set LCP of $N_l$ by $B_x$ |
| Until termination condition is met |

**Procedure:**

**Step 1** Obtain the approximate rate of information flow on each path which is related with the chromosome representation.

**Step 2** Verify if there is an unmarked LCP, which has at least two incoming links sending the same packet. If there is no unmarked LCP, end.

**Step 3** For the present LCP, among its incoming paths which have the same packet, ensure if there is at least one path which may obtain a different packet from those that the present LCP has already obtained. If there is no such path, mark the present LCP, and then go to the previous step.

**Step 4** For the present LCP, amongst its incoming paths which have the same packet, randomly pick a path which can convey new flow to the present LCP. Randomly allocate such a new flow to that path, alter the related gene in the chromosome and then terminate.

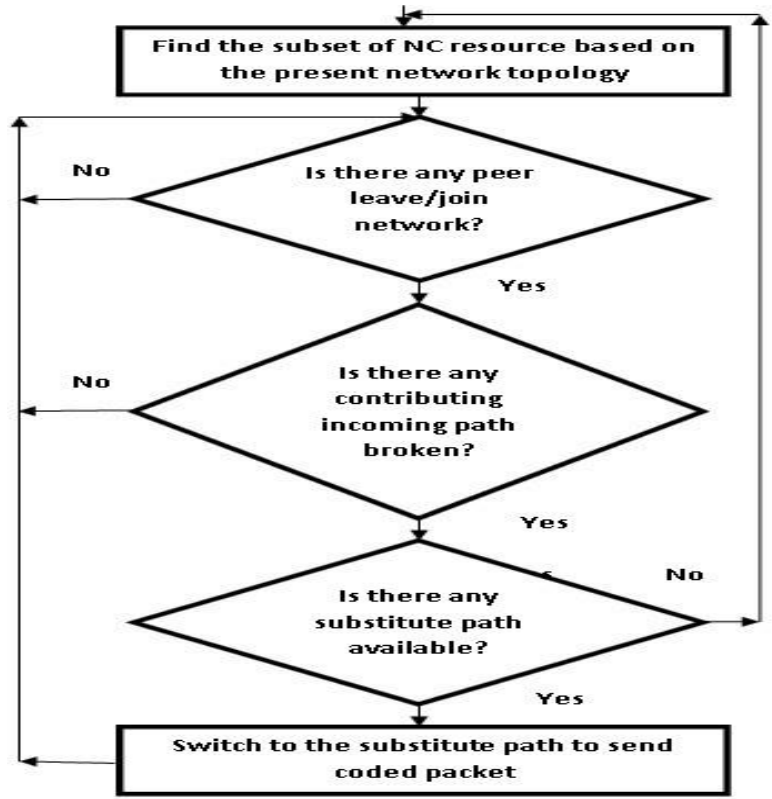

**Fig 3.** Proposed GA to the NCMP

### 5.3. Population initialization

Population initialization is a vital process in genetic algorithms since it influences the convergence quickness and also the quality of the optimum final solution[40]. If there is no information available about the solution, then random initialization is the commonly used technique to produce initial population. The success of Genetic Algorithm depends upon the individuals chosen in the initial population and the size of the population. If poor individuals are picked in the initial population, it will finish in longer execution time and weaker optimal solutions. Since the priori information about the solution to our problem is not clearly available, we use the concept of opposition based optimization techniques for the selection of the initial population. It helps to obtain near optimal starting candidate solutions.

The following steps are carried out for the opposition-based population initialization algorithm for the given problem which can be used instead of a random initialization:

(1) In the first step, uniformly distributed random population is generated, IP(s); s is the size of population;

(2) In second step, opposite population OP(s) is calculated. The $i^{th}$ equivalent opposite individual for OI(s) is calculated by

$$OI_{i,j} = c_j + d_j - P_{i,j}, \ i = 1, 2, \ldots, s; \ j = 1, 2, \ldots, N, \tag{6}$$

where N is the number of variables or dimension of the problem; $c_j$ and $d_j$ represents the interval boundaries of $j^{th}$ variable ($y_j \in [c_j, d_j]$);

(3) Finally, s fittest individuals from set the {IP(s) U OI(s)} are selected as initial population.

### 5.4. Parent Selection and Crossover scheme

Selection is the important step in a genetic algorithm in which individual genomes are selected from a population for later reproduction. An elitist model is approved in the proposed genetic algorithm for the selection of parents. The best individuals are selected and directly copy it to the next generation. We use roulette wheel selection mechanism to select the rest of individuals for each generation. The probability of parent selection $P_i$, represented by prob($P_i$), is given as

$$prob(P_i) = \frac{F(P_j)}{\sum_{k=1}^{s} F(P_k)} \tag{7}$$

The better individual has the higher probability of being picked as a parent according to the definition of fitness function. Hence, the common links between two parents represents the good individuality. Let $P_1$ and $P_2$ be the chosen parents. The crossover operator produces child $P_c$ by identifying the common link between $P_1$ and $P_2$, and retaining these links as same links in child $P_c$. We use the parameterized uniform crossover operator which indeed turns out to work better for our problem than conventional crossover operators. The correlations between coding points need not to be necessarily related to the proximity between their corresponding components in a chromosome. Therefore, parameterized uniform crossover will not cause any feasibility problem.

       **Procedure Roulette Wheel Selection**
       **Do**
       Set popl_size as 50
       Find cumulative fitness value, total fitness value (p) and sum of proportional fitness value (sum)
       Rotate the Wheel popl_size times

If sum<d then
Select the first chromosome, otherwise, select n[th] chromosome
End if
**Until** size of the population <popl_size
Return chromosome with fitness value proportional to the size of selected wheel section
**End**

## 5.5. Evaluation of chromosomes

Each chromosome is formed by either the initialization or generated by the evolutionary operators. It is assigned a fitness value that determines how well its corresponding solution matches to the problem, based on the optimization objective function of Eq. (1). Here, the chromosome's fitness value is defined as the corresponding collective actual throughput achieved. If this value is larger, then the chromosome is fit to the optimization problem.

The simple way to compute the fitness is to estimate fitness from limited trial. Suppose that the fitness gain is the mean of fitness for each of the trails. As these are calculated, the approximate of mean becomes correct. The fitness estimation is going to be utilized for picking the members of the population. It is good that this fitness can be calculated to be a useful level of accuracy long before the intact entourage of fitness cases has been evaluated. The estimate of fitness is the average of the individual fitness cases examined so far. Let us assume there are N of them, the trail mean is given by

$$F' = \frac{1}{N}\sum_{j=1}^{N} F_j$$

(8)

And the trail difference by $t^2 = \frac{1}{N-1}\sum_{j=1}^{N}(F_j - \bar{F})^2$

(9)

given that several independent estimate are going to be added, assume that the distribution of estimates of fitness is normal such as $D(\beta, \alpha)$. Then the trivial subsequent distribution of the mean is a scholar distribution with mean F', variance t/N, N-1 degree of freedom. A scholar distribution is the distribution of the random variable $v = \frac{(F'-\alpha)\sqrt{N-1}}{t}$, which is a method of estimating the mean of the normal distribution that does not depend on the variance. The collective mass function can be utilized to test for acceptance of the estimate. Once the trial mean is adequately close to the mean, examining fitness cases can be discontinued, and the estimate can be utilized in the reproduction phase.

## 5.6. Mutation operation and Termination criterion

When a new child is generated by applying the crossover operator, the mutation operation is done according to the mutation probability. Mutation adjusts one or more gene values in a chromosome from its early state. The solution may change completely from the previous solution in mutation. Therefore, GA can move toward to better solution by using mutation operator. Mutation takes place during evolution process according to a user defined mutation probability which should be set low. We use straightforward binary mutation, where each bit in each chromosome is flipped autonomously with probability 0.01. First, the mutation procedure randomly selects a subset of coding nodes and makes the nodes as non coding nodes in the network by removing all the links that connect these selected nodes and their outermost node on P. The new generation successively replaces the current one in the each iteration of evolutionary process. When a pre specified generation number is reached or when successive generations bring normalized improvements in actual throughput that is all smaller than a fixed threshold value, the proposed GA is terminated.

### 5.7. Generalization of proposed method

We use the standard structure of the GA introduced by Holland [13] with its elements particularly designed to suit our problem. Note that performance of GA depends on the specifications of its elements such as selection mechanism, crossover operation, mutation operation and numerical parameters, etc. Theoretical works, on the other hand, were not yet able to give a useful prediction about which combination of such elements will be best suited to a specific problem [11]. Hence we tested different selection of the elements reported to work well in several other studies, and chosen the one that worked well for our problem. Advantage of our proposed approach with compared to the random selection of nodes (RSN) and coding at all the nodes (CAN) approaches, it can be applied to a several range of generalized problems, which engage non-coding links and nodes and thus are difficult to resolve optimally.

**Number of Coding peers:** The proposed technique can generalize to the case of minimizing number of coding peers, which primarily was our objective.

**Cost of performing Coding operation:** If the coding cost is different at each of the links, it is necessary to minimize the total overhead needed by coding, which can be found by adding up the coding cost at each of the active coding points in the network and is used to calculate the fitness value of a reasonable chromosome. On the other hand, the previous approaches don't have a natural generalization to this situation unless the cost of coding can be clearly ordered, in which case pass through the links or peers in descending order of cost looks logical.

**Link Cost consideration:** The cost for usage of link is also evidently focus to optimization, which alone can be resolved efficiently by considering coding at all possible places while joint optimization over the coding and link costs is difficult. It is noted that the link cost optimization jointly with coding cost has been included into our GA-based approach by adjusting the fitness value as follows: represents the number of incoming and outgoing paths of peer, maximum possible throughput obtained in terms of download time and system throughput, actually achieved download rate to minimization of network coding resources has already been considered into account.

### 6. Experimental Result and Analysis

Since the code minimization in P2P network is a relatively new application of genetic algorithms, little work has been reported so far. The performance of proposed GA based node selection (GANS) is assessed by comparing the actually obtained network throughput in random selection of nodes (RSN) to perform coding operation and coding at all the nodes (CAN) in the network. In the experiments on the NCMP, there are two sets of test cases, taken from [22] and modified for relative analysis. The networks in NSet-I are generated by the algorithm in [38], which creates connected acyclic directed graphs randomly. Two networks such as NSetC1 and NSetC2 are considered and utilized for simulation in NSet-I. There are 30 nodes, 90 links, 20 receivers and rate 5 in NSetC1 and 50 nodes, 130 links, 30 receivers and rate 4 in NSetC2. **Table 3** shows the value of the various parameter's used in the proposed GA.

**GA parameters:**

| Parameters | Value |
|---|---|
| Population size | 50 |
| Crossover Probability | 0.8 |
| Mutation rate | 0.01 |
| Maximum Number of Generation | 100 |

**Table 3.** Parameters used in the proposed GA.

**Performance metrics:**

The following parameters are used for evaluating the performance of the proposed system:

1.  **Packet redundancy:** This metric is the ratio between the total amount of network traffic and smallest amount of network traffic in the ideal case. Ideally, only one copy of the original file flood in a network and the downloading peers can collect and then rebuild the original file through cooperation.
2.  **Average packet distribution time**: It is evaluated by the average time all the participant peers need to complete it's downloading of enough coded packet to construct the original file.
3.  **Maximum download time**: It is measured by the maximum amount of time for a peer to complete its download.
4.  **System throughput**: It is defined as the number of bytes reconstructed successfully by all the peers per second.
5.  **Failure rate**: It is referred as the number of peers unable to finish its download due to missing of some blocks.

The proposed mechanism performance in NSet-I and NSet-II is reported in **Fig. 5** and **Fig. 6**. **Fig. 5(a)** and **Fig. 6(a)** show the download time of the previously mentioned three different techniques in NSet-I and NSet-II for different file size. It is proved that GA based node selection reduces the download time with compared to other two different techniques in both network topology. Since optimal node selection is carried out, download time is reduced in GA based approach. **Fig. 5(b)** and **Fig. 6(b)** show how the proposed GA based approach resist against peer dynamics. In order to simulate this, we make some links as dynamic one. The randomization introduced by network coding makes each packet equally important. The coding operations are carried out and performed in the optimal nodes which make the system robust against link failures. **Fig. 5(c)** and **Fig. 6(c)** prove that the proposed system reduces packet redundancy with compared to other system even when file size increases. System throughput is increased and same in the first 400 minutes in all those three techniques. There is a slight improvement in GA based technique after 400 minutes in system throughput. It is reported in **Fig. 5(d)** and **Fig. 6(d).**

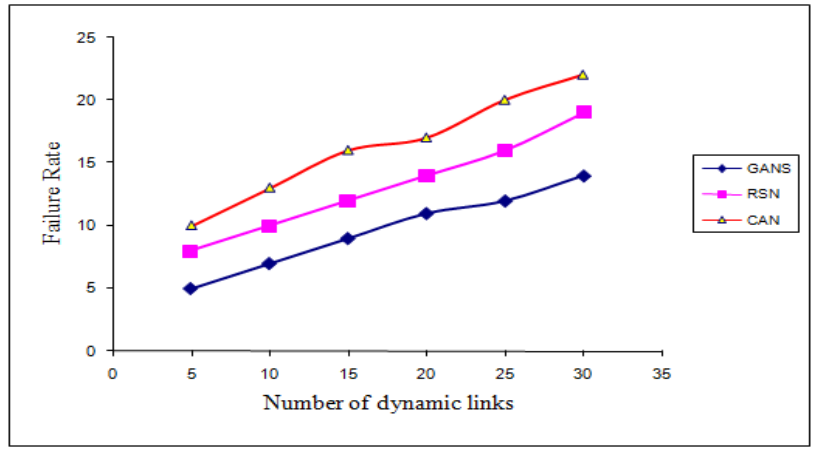

**Fig. 4.** a. Download Time Vs File Size

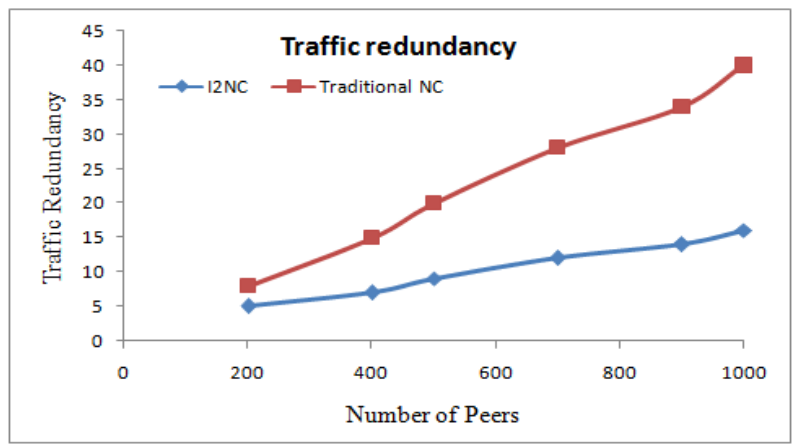

**Fig. 4.** b. Failure rate Vs Number of dynamic links

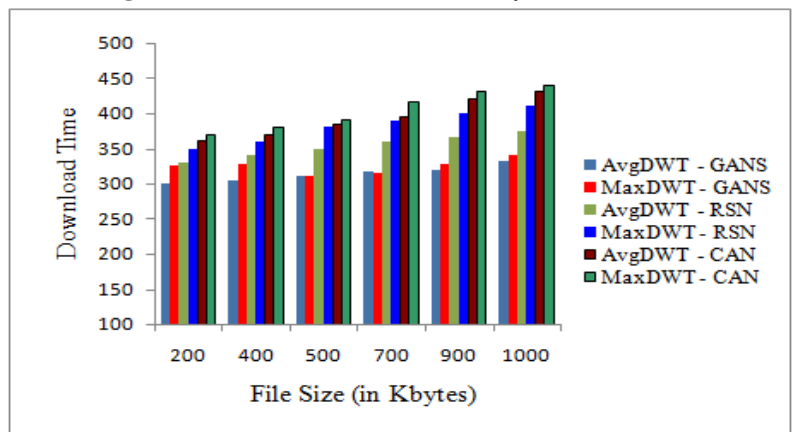

**Fig. 4.** c. Packet Redundancy Vs File Size

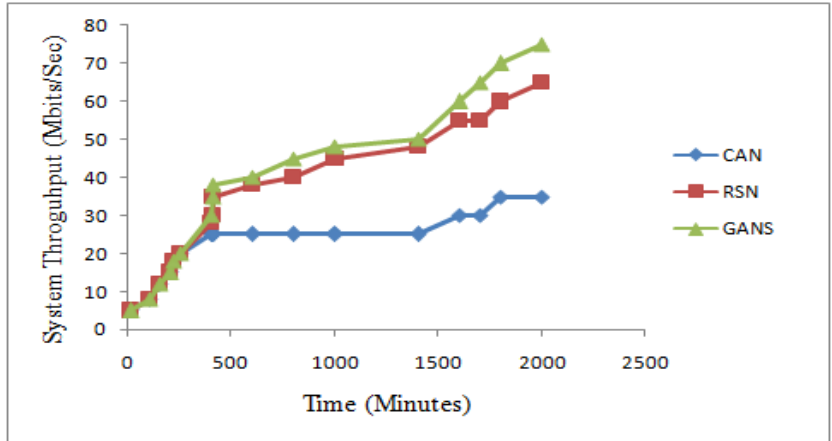

**Fig. 4.** d. System Throughput
**Fig. 4.** NSet - I – Performance analysis in NsetC1

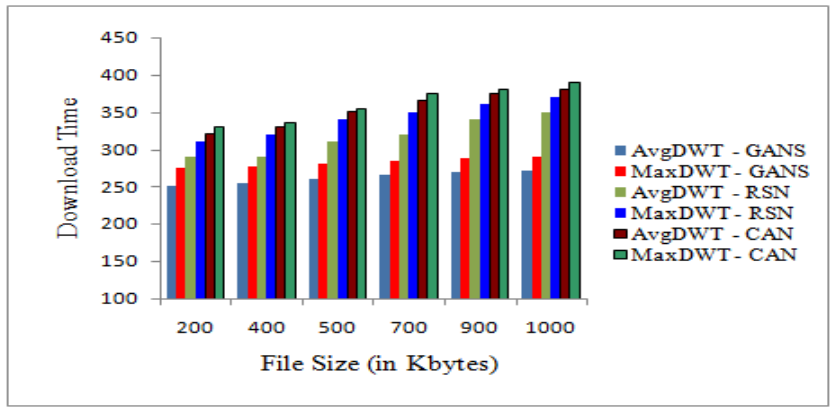

**Fig. 5.** a. Download Time in different runs

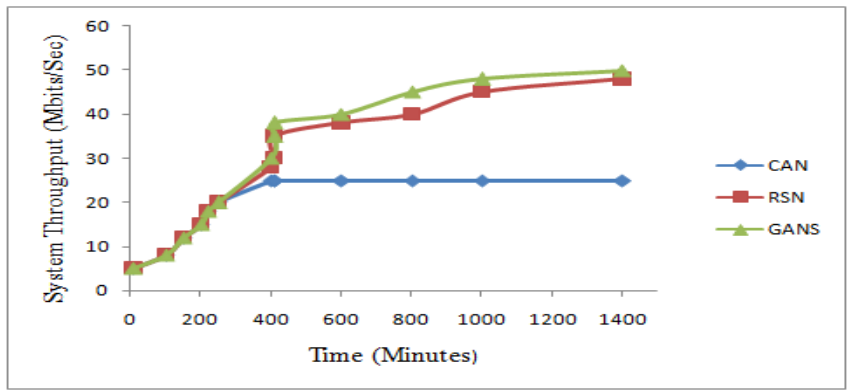

**Fig. 5.** b. Failure rate Vs Number of dynamic links

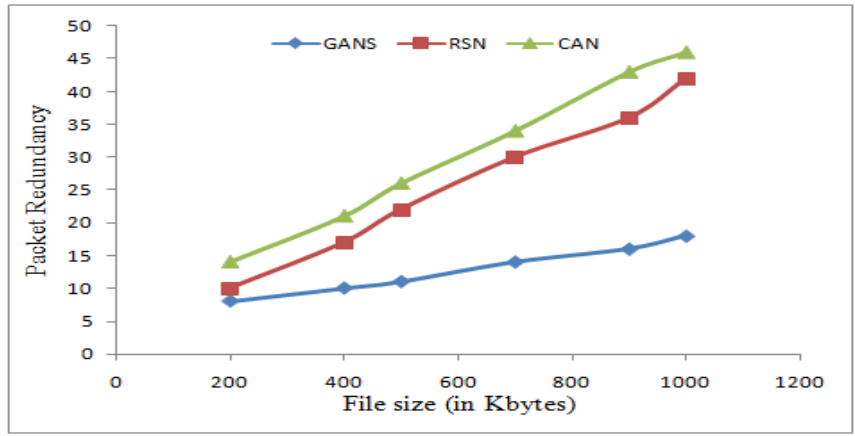

**Fig. 5.** c. Packet Redundancy Vs File Size

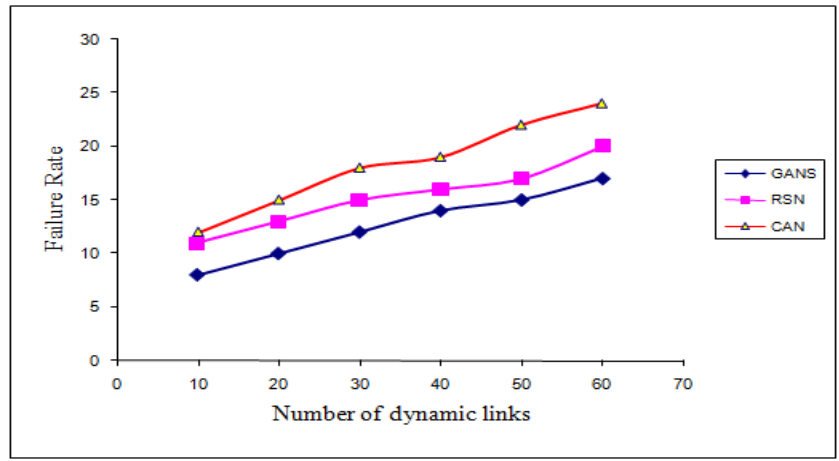

**Fig. 5.** d. System Throughput

**Fig. 5.** NSet - I – Performance analysis in NsetC2

## 7. Conclusion

Network coding is a proven technique to solve rarest block problem in P2P network and also it can be used in network protocol, wireless networks, and network security. It is necessary to decrease the coding operation carried out in network code based P2P content distribution due to the complexity involved. The optimization of network coding aims to reduce network coding resources such as coding nodes and links. It has recently attracted research attention. This paper considered how to address the network coding minimization problem (NCMP) by describing its general formulation, and then developing an efficient Genetic Algorithm (GA) to realize it. GA is used to find the optimal node to perform the coding operation in the network code based P2P network in order to achieve the optimization of merits and demerits of network coding. Its performance relies on its various elements such as initialization of population, crossover and mutation operation. In this proposed GA, we used problem specific GA operators such as opposition based technique for population initialization and parameterized uniform crossover operator. Both operators work better for our code minimization problem. The proposed method compared with two different techniques by focusing on mainly download time and redundant coded packet transmission. The simulation outcome proves that proposed method work better in terms of download time and packet redundancy rate as well as failure rate. For future work, multiple packet flow from different sources to multiple different receivers will be considered. The GA stated in this article may be improved since it considers the network topology and its scalability may be poor in large scale network. Hence, it will be improved by developing its distributed versions.